\documentclass[a4paper,fleqn,usenatbib]{mnras}

\usepackage[T1]{fontenc}
\usepackage{ae,aecompl}

\usepackage{graphicx}	
\usepackage{graphics}
\usepackage{amsmath}	
\usepackage{amssymb}	

\newcommand       \be           {\begin{equation}}
\newcommand       \ee           {\end{equation}}
\newcommand       \bea          {\begin{eqnarray}}
\newcommand       \eea          {\end{eqnarray}}

\title[Reconstructing the galaxy redshift distribution]{Reconstructing the 
galaxy redshift distribution from angular cross power spectra}

\author[Sun, Zhan and Tao]{Lei Sun$^{1,2}$\thanks{E-mail:
sunl@nao.cas.cn}, Hu Zhan$^{2}$ and Charling Tao$^{1}$
\\
$^{1}$Physics Department and Tsinghua Center for Astrophysics (THCA), Tsinghua University, Beijing, 100084, China\\
$^{2}$Key Laboratory of Space Astronomy and Technology, 
National Astronomical Observatories, Chinese Academy of Sciences, 
Beijing 100012, China
}

\date{Accepted XXX. Received YYY; in original form ZZZ}

\pubyear{2015}

\begin{document}
\label{firstpage}
\pagerange{\pageref{firstpage}--\pageref{lastpage}}
\maketitle

\begin{abstract}
The control of photometric redshift (photo-$z$) errors is a crucial and challenging task 
for precision weak lensing cosmology.  
The spacial cross-correlations (equivalently, the angular cross power spectra) of galaxies between tomographic photo-$z$ bins  
are sensitive to the true redshift distribution $n_i(z)$ of each bin and hence
can help calibrate the photo-$z$ error distribution for weak lensing surveys. 
Using Fisher matrix analysis, we investigate the contributions of various components of the angular power 
spectra to the constraints of $n_i(z)$ parameters and demonstrate the importance of the cross power spectra therein, 
especially when catastrophic photo-z errors are present. 
We further study the feasibility of reconstructing $n_i(z)$ from galaxy 
angular power spectra using Markov Chain Monte Carlo estimation.
Considering an LSST-like survey with $10$ photo-$z$ bins, we find that the underlying redshift  
distribution can be determined with a fractional precision ($\sigma(\theta)/\theta$ for parameter $\theta$) of roughly $1\%$ 
and $10\%$ for the mean redshift and width of $n_i(z)$, respectively. 
\end{abstract}




\section{Introduction}
\label{sec:intr}
Weak gravitational lensing is considered a powerful cosmological probe, and 
much work has been done in the past few decades to advance its measurement
technique, its application in cosmology, and our understanding of its 
systematics \citep[e.g.,][]{bacon01, heymans06, miller07, bridle09, hoekstra02, jarvis03, rhodes04, heymans04, hoekstra06, semboloni06, massey07, 
benjamin07, kilbinger13, jee13, liu14, fu14, Kohlinger15, fan07, yu15, joachimi15}.
With the success of precursor projects (e.g., CTIO, COSMOS, DLS, CFHTLenS),
several more ambitious weak lensing surveys
targeting at hundred to thousand times larger areas, such as
DES\footnote{http://www.darkenergysurvey.org},
LSST\footnote{http://www.lsst.org}, and
EUCLID\footnote{http://sci.esa.int/euclid/}, are ongoing or under construction.
Dark energy is a key science driver of these surveys.
In the meantime, various systematic effects (see reviews by \citealt{refregier03}, 
\citealt{bartelmann10} and \citealt{kilbinger15}) can become dominant contaminations as statistical 
errors are expected to reduce to insignificant levels for future surveys. 
The necessity of obtaining vast amount of galaxy redshifts in weak lensing observations makes the
multi-color photometric determination of redshifts the only practical means 
available.
The photometric redshift (photo-$z$) errors arising therefrom are among the most critical systematics. 
They are often characterized by the bias and scatter of the photo-$z$ $z_p$ with respect to 
the spectroscopic redshift (spectro-$z$) $z_s$, as well as the so-called
catastrophic errors, which can be loosely defined as the outliers with $|z_p-z_s|>1$.
Photo-$z$ errors contaminate the cosmological information in weak lensing and severely 
degrade the constraints on dark energy \citep[e.g.,][]{huterer06, ma06, zhan06, amara07, abdalla08, sun09, hearin10, hearin12}.
In order to fully exploit the potential of future weak lensing surveys like the LSST, 
uncertainties on both the bias and the scatter of $z_{p}$ have to
be controlled to below $0.003$ \citep[e.g.,][]{huterer06, ma06}, which is a stringent 
requirement. 

Recently, a technique is developed to calibrate the true redshift distribution 
of the photo-$z$ sample via cross-correlations with a spectro-$z$ sample in 
the same volume \citep{lima08, cunha09, matthews10, mcquinn13, schmidt13, menard13, rahman15, eriksen15}. 
The spectro-$z$ sample serves as an accurate redshift reference. As one 
slices through redshift space, the cross-correlation between the spectro-$z$ 
sub-sample around $z$ and the photo-$z$ sample increases with the overlap 
between the two and thus maps the true redshift distribution of the 
photo-$z$ sample. 
It is still challenging, however, to obtain a good sample of galaxy spectra
covering a sufficiently large area to moderately high redshift ($z\sim 4$) for
deep surveys like LSST. Moreover, such a calibration is normally valid
only in the volume where the photo-$z$ and spectro-$z$ samples overlap. To 
extend over the whole photo-$z$ survey area, one has to assume that the 
difference between the photo-$z$ sample inside and that outside the 
calibration area is negligible.

Separately, it is noted that the galaxy angular auto power spectra can provide
useful consistency test on the photo-$z$ error distribution \citep{zhan06b}.
\citet{zhan06} further shows with the simple parametrization\footnote{It is still 
considerably more flexible than a mean redshift and an rms error per photo-$z$ 
bin.} of a photo-$z$ bias and an rms error per redshift interval (not per bin) 
$\Delta z= 0.1$ that the photo-$z$ error distribution can be calibrated to 
satisfactory precision by the auto and cross power spectra of the very same 
galaxies in the weak lensing survey. As a result, the joint analysis of galaxy clustering 
(the galaxy angular power spectra) and weak lensing is much less prone
to uncertainties in the photo-$z$ bias and rms parameters. 
\citet{schneider06} also carries out a study, using a Fisher Matrix analysis, on how 
well the redshift distribution of galaxies in 6 bins can be determined by 
angular power spectra of the 6 bins alone. The results show that if the 
galaxies in each bin are allowed to be wrongly assigned to all other bins 
with no restriction, the degrees of freedom in the photo-$z$ error 
distribution would overwhelm the constraining power of the galaxy angular 
power spectra. Therefore, one should model the photo-$z$ error distribution 
with as few parameters as possible at an acceptable accuracy.
\citet{zhang10} combines the angular power spectra and weak lensing analysis to perform 
the self-calibration, based on Fisher matrix analysis as well.
They find that lensing-galaxy correlations also help to improve the photo-z self-calibration 
by breaking the degeneracy between up scatters and down scatters of photo-z samples.  
\citet{benjamin10} proposes to parameterize the redshift distributions with the total 
sample number $N_i$ and the contamination fraction from the $i$th to the $j$th bin as $f_{ij}$. 
With this parametrization, the authors investigate the feasibility of reconstructing $N_i$ 
and $f_{ij}$ of each bin in a 4-bin case both from simulated and real data. 
  
In this paper, we focus on the reconstruction of the true redshift distribution of 
each photo-z bin in tomographic weak lensing analysis, using angular cross power spectra. 
We first build a simple double-Gaussian model to characterize the true redshift 
distribution of galaxies within tomographic photo-z bins,  which is motivated by 
the bimodal distribution of photo-z errors in presence of catastrophic errors. 
This model can also be extended straightforwardly to a multi-Gaussian form 
in the case of multiple catastrophic fractions. 
Based on this model, we analyze the contributions of various components of 
the galaxy angular power spectra, such as the cross power spectra, 
the baryon acoustic oscillations (BAO) and the broadband shape, 
to the reconstruction of the underlying redshift distribution.
The performance of this calibration method in practical data analyses is 
investigated with 10 tomographic photo-z bins, utilizing simulated data and a 
Markov Chain Monte Carlo(MCMC) estimate \citep[e.g.,][]{gilks96}.  
The simple double-Gaussian model suffices the purpose of this work, i.e., 
demonstrating the feasibility of this self-calibration method. The precise modeling 
of the galaxy true z-distribution in a given photo-z bin, however,  will be crucial 
for the control of photo-z errors in future weak lensing surveys. Thus we plan 
to perform directly a principle component analysis (PCA) to characterize the 
probability distribution function $P(z_s|z_p)$.

The rest of this paper is arranged as follows. 
In \S 2, the galaxy angular power spectra and the redshift distribution model 
used in our study is introduced. 
In \S 3, we analyze the contributions of various components on galaxy angular power 
spectra to the reconstruction of the true redshift distribution. 
Using MCMC estimate, we study the feasibility of this reconstruction method in 
practical analysis in \S 4.  A summary is presented in \S 5.

\section{Method}
\label{sec:method}

We first describe how the galaxy angular power spectra is calculated theoretically in our study.   
We then present the (double-)Gaussian model that is used to characterize 
the galaxy true redshift distribution in each tomographic photo-z bin. 

\subsection{Galaxy Angular Power Spectra}
\label{sec:pl}

The galaxy angular power spectrum, under the Limber approximation(Limber 1954), can be written as (Hu \& Jain 2004) 
\be 
\label{eq:pl}
P_{ij}(\ell) = \frac{2\pi^2}{c\ell^3} \int \rm dz\, H(z) D_{\rm A}(z) W_i(z) W_j(z) \Delta^2_m(k; z),
\ee
\noindent where the subscripts $i,j$ label the tomographic bins. $H(z)$ and $D_A(z)$ are the Hubble parameter
and comoving angular diameter distance, respectively. 
$\Delta^2_m(k; z)$ denotes the dimensionless matter power spectrum, where $k=\ell/D_A(z)$ is the wavenumber
 that projects onto multiple $\ell$ at redshift $z$. $W_i(z)$ is the 
weighting function, $W_i(z)=b(z)n_i(z)$, with $b(z)$ the linear galaxy bias 
and $n_i(z)$ the normalized (to $1$)
galaxy redshift distribution of the $i$th tomographic bin (hereafter denoted as bin-$i$). 
We discuss the modeling of $n_i(z)$ specifically in Section \ref{sec:model}.

The nonlinear matter power spectrum can be obtained by: 
\be
\label{eq:nl}
P_{\rm nl}=(P_{\rm lin}-P_{\rm nw})\exp(-\frac{k^2}{2k_{\star}^2})+P_{\rm nl,nw},
\ee
\noindent where the first half processes damping of the BAO signal because of 
nonlinear evolution, with $k_{\star}$ being the characteristic damping scale \citep{eisenstein07}. 
$P_{\rm lin}$, $P_{\rm nw}$ and $P_{\rm nl,nw}$ are 
the linear matter power spectrum, its no-wiggle approximation \citep{eisenstein98} and 
the nonlinear corrected no-wiggle power spectrum, respectively. 

Throughout this study, we consider a fiducial survey resembling the LSST design, 
with $20,000$ square degrees 
and a redshift range of $0 \leq z \leq 4$.
Following \citet{Zhan06a}, we model the overall redshift evolution of the galaxy bias as $b(z)=1+0.84z$. 
Within each bin the galaxy bias is a constant and its fiducial value 
is assumed at the mean redshift $z_m$ of the individual bin. 
The galaxy bias parameters then float freely (not bound to the floating $z_m$), 
and will finally be marginalized over in the following specific 
calculations, with a relative $20\%$ prior.

We discard information from very large scales, $\ell < 40$, uniformly for 
all tomographic bins to avoid potential large-scale effects that are not 
included in this study (e.g., dark energy clustering). 
A high multipole cut for each galaxy bin is also 
incorporated to reduce contaminations of the small-scale non-linearity and 
baryonic effects. According to \citet{zhan06}, the maximum multipole for a galaxy bin centred at $z_{\rm m}$ 
follows approximately 
$\ell_{\rm max} =340z_{\rm m} + 346z_{\rm m}^2$ at
$0.15 \le z_{\rm m} \le 0.9$, 
$446 - 658z_{\rm m} + 908z_{\rm m}^2$ at 
$0.9 < z_{\rm m} \le 2.1$, and 3000 at $z_{\rm m} > 2.1$.
Roughly speaking, we make use of the information of galaxy angular spectra on scales $40 \le \ell \le 3000$.

The redshift distortion (RSD) effect is not modeled here. It is largely erased  
in the angular power spectra due to projection along the line of sight, 
provided that the photo-$z$ bins are wide enough in redshift. 
In addition, the cuts at low(high)-$\ell$ ends mentioned above should also help in reducing the impact of   
the linear RSD on large scales and the 'Finger of God' effect on small scales, respectively.

We consider a flat $\Lambda$CDM model with cosmological parameters 
$w=-1, \Omega_m=0.27$, $\Omega_b=0.0446$, $n_s=0.96$, $h=0.72$ and $\sigma_8=0.78$. 
Throughout, the cosmological parameters are fixed at their fiducial values. 
The ultimate goal of the self-calibration scenario is to
combine the weak lensing and galaxy clustering analysis together 
to constrain both cosmological and redshift distribution parameters simultaneously.
With angular power spectrum alone, here we shall focus on its performance 
in constraining redshift distribution parameters.

\subsection{Modeling the galaxy redshift distribution}
\label{sec:model}

For future large sky weak lensing surveys, precise reconstruction of the source 
galaxy redshift distribution from the 
defective photo-z measurements is a critical and challenging task. 
Lots of efforts are made to better model the true redshift distribution $n_i(z)$, 
given a tomographic photo-z bin $i$.
In previous studies of photo-z calibration using cross correlation techniques, 
$n_i(z)$ is usually parametrized by a total galaxy number $N_i$ with catastrophic 
contamination fractions $f_{ij, j\neq i}$, i.e., 
fraction of galaxies mis-assigned to bin-$i$ from bin-$j$ \citep[e.g.,][]{schneider06, benjamin10, zhang10}. 
In this case, $n_i(z)$ is shaped as a histogram. With finer binning within bin-$i$ 
(correspondingly more parameters), a smooth curve of $n_i(z)$ can also be expected. 

This work emphasizes effectiveness and feasibility of the self-calibration method,
and we simply adopt a double-Gaussian to model $n_i(z)$, with one Gaussian for 
the main sample and the other for a sub-sample due to catastrophic failures. 
This model is inspired by the fact that the confusion of the $4000\AA$ 
break with the Lyman break in galaxy SEDs, the dominant source of catastrophic failures, 
often leads to a true redshift distribution of bi-model pattern in the relevant 
tomographic photo-z bins. If there exist various catastrophic failure fractions within one bin, 
this model can also be extended straightforwardly to a multi-Gaussian form. 

Specifically, the double-Gaussian distribution of bin-$i$ can be written as: 
\be
\label{eq:nz}
n_i(z)=(1-f^c)\exp{\frac{(z-z_{mi})^2}{\sigma^2_i}}+f^c\exp\frac{(z-z_{mi}^c)^2}{{\sigma_i^c}^2},
\ee
\noindent where $f^c$ is the fraction of catastrophic failures with respect to 
the total sample of bin-$i$. $z_{mi}$ and $\sigma_i$ denote the mean redshift and 
the width of the distribution of the main sample, whereas $z_{mi}^c$ and $\sigma_i^c$ 
are analogues for the sub-sample of catastrophic failure. 

\section{Contributions of various components of the angular power spectra}
\label{sec:fisher}

Using Fisher matrix analysis, we clarify the contributions of various components 
of the galaxy angular power spectra, such as cross(auto) power spectra and BAO, 
to the constraints of galaxy redshift distribution parameters. 

\subsection{Parameters}
\label{sec:fish-params}

As mentioned in Section \ref{sec:pl}, our fiducial survey is a LSST-like half sky survey 
with a redshift range $0 \leq z \leq 4$. 
We consider dividing the whole redshift range into 5 tomographic bins for clarity here, 
whereas thinner binning can be performed in practical analyses.

\begin{figure}
\includegraphics[width=\columnwidth]{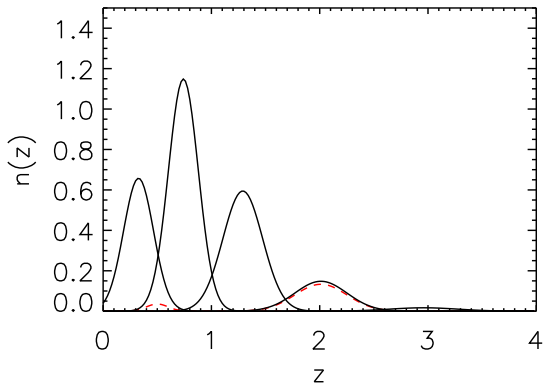}
 \caption{Galaxy redshift distribution curves of the 5 bins. 
Black solid lines denote those with no catastrophic case. 
The corresponding $z_{mi},\sigma_i$s are given in Table \ref{tab:params}. 
The 2 red dashed lines present the true distribution of galaxies assigned to bin-$4$, 
when there is a $10\%$ contamination of catastrophic failures from bin-$1$. }
\label{fig:nz-cata}
\end{figure}

As a first step, we examine the case free of catastrophic contaminations.
The specific values of redshift distribution parameters $z_{mi},\sigma_i$ of each bin 
are listed in Table \ref{tab:params}, 
which are determined from the overall galaxy redshift distribution in \citet{zhan06}. 
The black solid lines in Fig. \ref{fig:nz-cata} illustrate the corresponding distribution functions $n_i(z),i=1,...,5$, 
each normalized according to the proportion of their surface number density $n_g^i$ to the total surface 
number density $n_g^{\rm tot}$. 
Here, the total surface number density of sampled galaxies is taken to 
be $40$ arcmin$^{-2}$, and $n_g^i$ of each individual bin (see the $5$th column of Table \ref{tab:params}.) 
is set by integrating the radial selection function of LSST \citep{zhan06}.
Note that the adjacent bins are set overlapped in our model, which naturally mimics 
the underlying distributions of practical photo-z samples to a certain extent. 
The cross power spectra are determined by the overlap between bins in true redshift space.   
   
We next include the catastrophic photo-z errors by considering that bin-$4$ 
suffers a fraction of $f^c=10\%$ contamination from bin-$1$.
The underlying distribution of $n_4(z)$ is thus replaced by the double-Gaussian model, 
shown as the combination of the two red dash lines in Fig. \ref{fig:nz-cata}. 
$z_{m4}^c=0.5$ and $\sigma_4^c=0.1$ are the mean redshift and width of the catastrophic sub-sample, respectively.  
The parameters of the main sample of bin-$4$, $(z_{m4},\sigma_{4})$, 
are kept unchanged (only normalization changed to $1-f^c$).

\begin{table}
\caption[]{Redshift distribution and survey related parameters. $z_m$, $\sigma$, 
$\ell_{max}$ and $n_g$ are the mean redshift , width, maximum $\ell$ and galaxy number density of each bin.}
\label{tab:params}
 \begin{center} \begin{tabular}{clclclclcl}
 \hline\noalign{\smallskip}
 \hline\noalign{\smallskip}
 bin &$z_{m}$  & $\sigma$   & $\rm \ell_{max}$ & $\rm n_g /arcmin^2$  \\
  \hline\noalign{\smallskip}   
 1   & 0.33 & 0.14       & 136       &   8.92               \\
 2   & 0.74 & 0.14       & 425       &  15.58     \\
 3   & 1.29 & 0.18       & 1081      &  10.69     \\ 
 4   & 2.01 & 0.24       & 2785      &   3.50     \\ 
 5   & 2.96 & 0.32       & 3000      &   0.53                \\
  \noalign{\smallskip}\hline 
  \end{tabular}\end{center}
\end{table}

\subsection{The Fisher Matrix}
\label{fish-fish}

The Fisher matrix takes the form (\citealt{tegmark97}; for this derived form see also explanations in \citealt{zhan06})
\be
F_{ij} = \sum_{\ell} \,
\left ({\partial {C_{ij}(\ell)}\over \partial \theta_i}\right )^T\,
{\rm Cov}^{-1}_{ij,mn}(\ell,\ell')\,
{\partial { C_{mn}(\ell')}\over \partial \theta_j},
\label{eq:fish}
\ee
\noindent where the total angular power spectrum $C_{ij}(\ell)$ is given by
\be
C_{ij}(\ell)=P_{ij}(\ell) +\delta_{ij} {1 \over n_g^i}.
\label{eq:c_obs}
\ee
\noindent ${\rm Cov}^{-1}_{ij,mn}(\ell,\ell')$ is the inverse of the covariance
matrix. Considering only shot noise and the Gaussian sample variance,
it can be written as
\bea
&{\rm Cov}\left [C_{ij}(\ell), C_{mn}(\ell')\right ] = {\delta_{\ell \ell'}\over (2\ell+1)\,f_{\rm sky}\,\Delta \ell}\, \nonumber \\
&\times \left [ C_{im}(\ell) C_{jn}(\ell) + C_{in}(\ell) C_{jm}(\ell)\right ]. 
\label{eq:cov}
\eea
\noindent Here $\Delta \ell=1$ is the band width, and $f_{\rm sky}\approx0.485$
is the fractional sky coverage of the fiducial survey. 

Beyond the cross power spectra, we are also interested in the role of BAO, a 
minute feature in the power spectrum that is used as a standard ruler to 
measure angular diameter distances, in constraining $n(z)$.
We thus consider 3 cases in which the power spectrum in equation
(\ref{eq:c_obs}) is modified as follows: 1) the 'auto-P' case with 
cross power spectra abandoned; 2) the 'no-wiggle' case with BAO wiggles 
edited out; 3) the 'only-wiggle' case with the broadband shape of the power 
spectra divided out. 

The 'auto-P' case is simply obtained by keeping only the auto power spectra within each z-bin 
and removing completely the cross correlation information between bins. 
A comparison with the 'full-P' case, which includes the full information of tomographic angular power spectra, 
can thus demonstrate clearly the role of cross power spectra.
In the 'no-wiggle' case, the Fisher matrix is computed using the nonlinear corrected 
no-wiggle matter power spectrum ($P_{nl,nw}$, see Section \ref{sec:pl}). 
In the 'only-wiggle' case, the broadband shape of the power spectrum is fitted with 
a $3$rd order smooth polynomial and then divided out, whose dependence on $n(z)$ parameters 
is consequently removed from the Fisher matrix. By comparisons of the 'no(only)-wiggle' 
cases versus the 'full-P' case, the effect of BAO can then be isolated. 

\subsection{Results}
\label{sec:fish-results}

\begin{figure}
\includegraphics[width=\columnwidth]{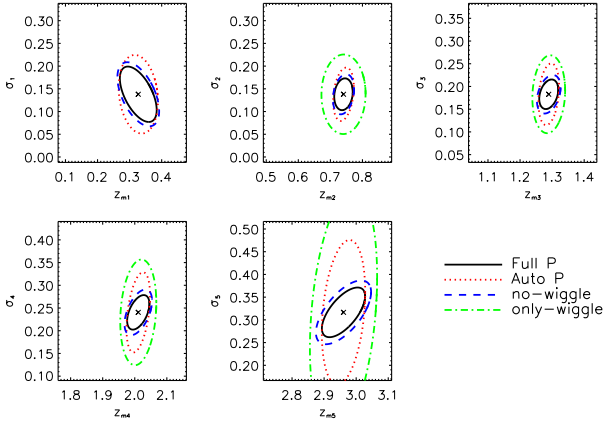}
 \caption{The redshift distribution parameter constraints from various components of 
the galaxy angular power spectra in the 'no-cata' case. Contours denote the $1-\sigma$ credible regions.}
\label{fig:fish}
\end{figure}

\begin{figure}
\includegraphics[width=\columnwidth]{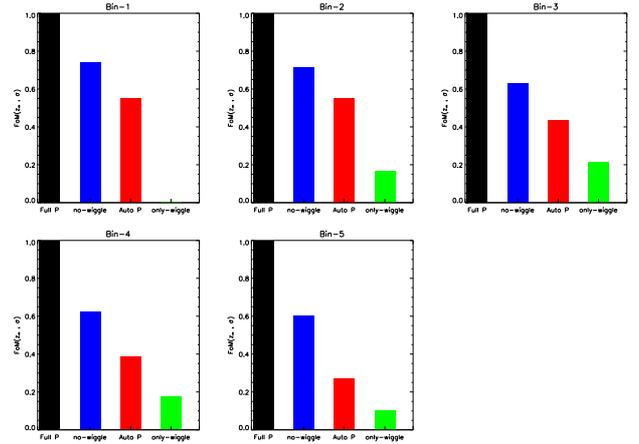}
 \caption{Relative F.o.M of $z_{mi}-\sigma_i$ for each bin in the 'no-cata' case, 
with the Full-P results normalized to 1. }
\label{fig:fom}
\end{figure}

We first derive constraints on the redshift distribution parameters when the catastrophic 
failure fractions are ignored (the 'no-cata' case).  Fig. \ref{fig:fish} shows the $1$-$\sigma$ 
error contour of $z_{mi}$-$\sigma_i$ of each bin for each considered situation, 
with a fractional $20\%$ prior applied on galaxy bias $b_i$ and the other cosmological parameters 
fixed at fiducial values. The $2$-D contour is then compressed to a Figure of Merit 
(F.o.M)\footnote{F.o.M = $\sqrt {\rm det (F)}$, where F is the reduced Fisher matrix 
for $z_{mi}$ and $\sigma_i$.} in Fig. \ref{fig:fom} for a quantitative comparison.  
 
By comparing the 'auto-P'(red dotted lines) with the 'full-P'(black solid lines) results 
from Fig. \ref{fig:fish}, a significant enlargement of the error region is noticed for all 5 bins 
when the cross power spectra are discarded. Furthermore, the bin width parameters $\sigma_i$ 
rather than the mean redshift $z_{mi}$ gain mostly from the inclusion of the cross power spectra. 
This implies that the overlapping of adjacent bins caused by wider tails dominates 
the cross correlation signals. Fig. \ref{fig:fom} shows that the omission of the cross power spectra 
(red bars) can generally result in $45\%$-$75\%$ degradation of the F.o.M of $z_{mi}$-$\sigma_i$ 
for different bins. It is confirmed that the cross power spectra play a crucial role in constraining 
the parameters of redshift distributions.

The BAO feature is also helpful in constraining the redshift distribution parameters, 
unveiled by the contrast of the 'no-wiggle' with the 'full-P' cases. 
The blue dashed contours in Fig. \ref{fig:fish}, which are slightly loosened on 
both dimensions compared to the solid ones, indicate that the BAO wiggles contribute almost 
evenly to constraints of $z_{mi}$ and $\sigma_i$. The capability of angular (cross) power spectra 
in self-calibrating the redshift distributions would be ruined if it was a strict power law 
in $\ell$ due to degeneracies between the $n_i(z)$ parameters and the normalizations \citep{zhang10}. 
Hence the BAO wiggles help in this regard. Specifically, for bins centered at different $z_m$ and/or 
with different widths, the BAO wiggles are projected at different $\ell$. Removing the BAO wiggles 
gives rise to a $\sim30$-$40\%$ degradation of the F.o.M of $z_{mi}$-$\sigma_i$ as shown in 
Fig. \ref{fig:fom}, though it affects less than removing the cross power spectra does. 

Discarding the broadband shape of the power spectra, i.e., the 'only-wiggle' case 
(green dash-dotted lines in Fig.\ref{fig:fish} and green bars in Fig.\ref{fig:fom}), 
induces a big loss of constraining power.  The F.o.M in this case decreases more than $80\%$. 
The inconsistency between 'no-wiggle' and 'only-wiggle' results is mainly 
due to the fact that the division operation in the 'only-wiggle' case induces 
a further loss of the absolute amplitude of the BAO.
Note that due to the high-$\ell$ cut at $\ell<136$, there is nearly no BAO signal counted in bin-$1$, 
therefore no constraints reported from it for the 'only-wiggle' case.

\begin{figure}
\includegraphics[width=\columnwidth]{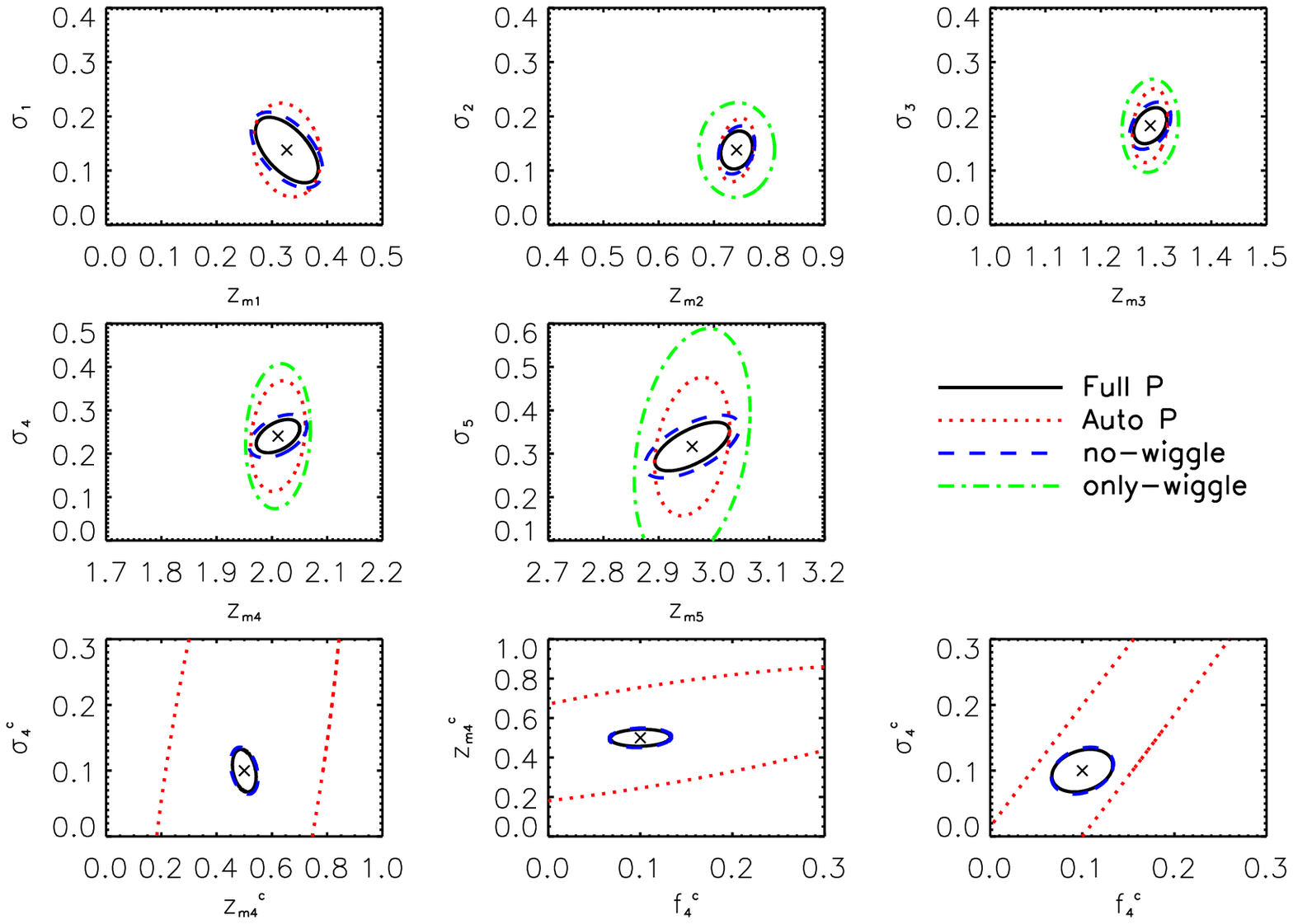}
 \caption{The redshift distribution parameter constraints from various components of 
the galaxy angular power spectra in the 'with-cata' case. Contours denote the $1-\sigma$ credible regions.}
\label{fig:fish-cata}
\end{figure}

\begin{figure}
\includegraphics[width=\columnwidth]{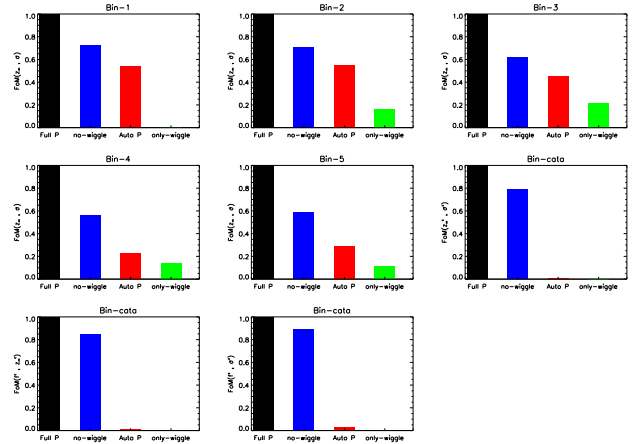}
 \caption{The same as Fig.~\ref{fig:fom} but for the catastrophic case.}
\label{fig:fom-cata}
\end{figure}

Next we consider the case with the assumed $10\%$ catastrophic fraction in bin-$4$ (the 'with-cata' case).  
The $1$-$\sigma$ error region of $z_{mi}$-$\sigma_i$ of each bin and the corresponding F.o.Ms 
are shown in Fig. \ref{fig:fish-cata} and \ref{fig:fom-cata}, respectively.
First of all, the constraints on $z_{mi}$-$\sigma_i$ of the main sample of bin-$4$ 
and of the other bins are not affected significantly by the considered catastrophic contamination. 
In the 'full-P' case for instance, the F.o.M of the 'with-cata' case shows a $\sim 15\%$ decrease 
for bin-$4$ main sample and $\leq 5\%$ for the remaining  bins, compared to that of the 'no-cata' case. 
Concerning the catastrophic fraction related parameters $f^c, z^c$ and $\sigma^c$, 
the constraints are undoubtedly sensitive to whether the cross power spectra are included or not, 
since the catastrophic contamination is the main source of the cross correlation between distinct bins. 
As a result, the 'auto-P' case gives nearly no constraints on these three parameters, 
with corresponding F.o.Ms $\to 0$. Besides, as the contamination is originally from the lowest bin-$1$, 
the absence of BAO wiggles in the cross correlation between bin-$cata$ and bin-$1$ 
nullifies the 'only-wiggle' case in constraining $f^c, z^c$ and $\sigma^c$. 

\section{Feasibility of reconstructing true redshift distributions using angular power spectra} 
\label{sec:mcmc}

So far the analysis is based on Fisher matrix, namely, theoretical calculations around the fiducial value. 
In this section we focus on MCMC estimates of the underlying 
redshift distribution parameters using simulated angular power spectra data, 
aiming to investigate the performance of this method in a practical data analysis. 

\subsection{Parameters}
\label{sec:mcmc-params}

We consider a more realistic case with 10 tomographic redshift bins, among which bin-$1$ and $7$ 
are contaminated by catastrophic errors. 
The two catastrophic error parts are chosen based on the LSST photo-z simulations, 
which shows two prominent clumps around the corresponding positions in the $z_p-z_s$ plot (Fig. 3.16, \citealt{LSST09}). 
While information extracted from the weak lensing analysis tend to be saturated more rapidly 
with increasing bin number due to the broad kernels, 
thinner binning usually does help the $2$-D galaxy clustering analysis (e.g, angular power spectra) 
to recover more $3$-D information along the line of sight, 
and consequently a more precise redshift distribution. 
Certainly, thinner binning means more free parameters, hence a balance has to be found for specific samples.    

\begin{figure}
\includegraphics[width=\columnwidth]{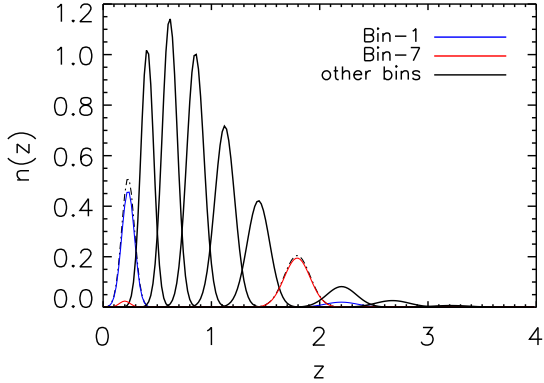}
 \caption{Galaxy distribution curves of the 10 bins in true redshift space. 
Black solid lines represent bins without catastrophic errors. 
The corresponding $z_{mi},\sigma_i$s are given in Table \ref{tab:cata2-params}. 
The blue(red) lines denote the true distributions of galaxies assigned to bin-$1$($7$), 
when there is a $10(5)\%$ contamination of catastrophic failures, respectively.
}
\label{fig:nz-2cata}
\end{figure}

The specific distributions of each bin in true redshift space are shown in Fig. \ref{fig:nz-2cata}. 
The combination of two blue solid lines denotes the distribution of bin-$1$, 
including its main sample at $z_m\sim0.2$ and $f^c=10\%$ catastrophic fraction around redshift $2.2$ (denoted as bin-$1^c$). 
Similarly, the two red solid lines show the distribution of bin-$7$, 
with main sample at $z_m\sim1.8$ and $5\%$ catastrophic fraction around redshift $0.2$ (denoted as bin-$7^c$).
The black solid lines represent distributions of the other bins. The corresponding parameters of each bin, 
listed in Table \ref{tab:cata2-params}, are regulated according to the same rules in Section \ref{sec:fish-params}.
     
\begin{table}
\caption[]{Redshift distribution and survey related parameters for the 10-bin case. 
$z_m$, $\sigma$, $\ell_{max}$ and $n_g$ are the mean redshift , width, maximum $\ell$ 
and galaxy number density of each bin.  bin-$1^c$ and $7^c$ represents 
the catastrophic sub-samples of bin-$1$ and $7$, respectively, 
and their fractions within each corresponding bin are given by $f^c$.}
\label{tab:cata2-params}
 \begin{center} \begin{tabular}{clclclclclcl}
 \hline\noalign{\smallskip}
 \hline\noalign{\smallskip}
 bin &$z_{m}$  & $\sigma$   & $\rm \ell_{max}$ & $\rm n_g /arcmin^2$  & $f^c$ \\
  \hline\noalign{\smallskip}   
 1      & 0.23 & 0.06       & 90       &   2.98     &  $10\%$  \\
 $1^c$  & 2.20  & 0.15     & --       & --          & --      \\
 2      & 0.41 & 0.06       & 190      &   5.95     &  --      \\
 3      & 0.61 & 0.07       & 330      &   7.77     &  --       \\
 4      & 0.85 & 0.08       & 540      &   7.82     & --       \\
 5      & 1.12 & 0.09       & 860      &   6.39     & --      \\
 6      & 1.43 & 0.10       & 1350     &   4.31     & --       \\
 7      & 1.79 & 0.12       & 2190     &   2.40     & $5\%$     \\
 $7^c$  & 0.20 & 0.05      & --       &   --        &  --       \\
 8      & 2.20 & 0.14       & 3000     &   1.10     & --        \\
 9      & 2.67 & 0.16       & 3000     &   0.41     & --       \\
10      & 3.21 & 0.18       & 3000     &   0.12     & --       \\
  \noalign{\smallskip}\hline 
  \end{tabular}\end{center}
\end{table}

\subsection{MCMC estimate and the Likelihood Function}
\label{sec:mcmc-like}

The MCMC method estimates parameters through mapping of the posterior distributions. According to Bayes' 
Theorem, one can infer ${\cal P}(\theta_i \mid \hat{P}_{ij})$, the posterior probability of 
parameters $\theta_i$ (e.g., $z_{mi}$ and $\sigma_i$), given measured galaxy 
angular power spectra $\hat{P}_{ij}$ from the likelihood function, once the prior ${\cal P}(\bf{\theta})$ is specified:
\be
{\cal P}( {\bf \theta} \mid \hat{P}_{ij}) = {\cal L}(\hat{P}_{ij} \mid P_{ij}({\bf \theta})){\cal P}(\bf{\theta}),
\label{eq:bayes}
\ee 

Concerning the likelihood, the correlation between angular power spectra of different redshift bins 
in a tomographic case gives rise to a multi-variate Gamma distribution, 
with a parameter-dependent covariance matrix. Its exact form is computationally unfeasible 
in practical estimation. A good approximate likelihood is therefore required. 
Based on discussion in \citet{sun13}, we choose to use the approximation of multi-variate 
Gaussian likelihood with its determinant term discarded,
\be
\ln {\cal L}\propto \left [P_{ij}(\ell)-\hat{P}_{ij}(\ell)\right ]{\rm Cov}^{-1}_{ij,mn}(\ell,\ell')
\left [P_{mn}(\ell')-\hat{P}_{mn}(\ell')\right ]\;.
\label{eq:coeff}
\ee

With cosmological parameters fixed at their fiducial values, it amounts to $38$ dimensions 
in parameter space to explore in the MCMC analysis, i.e., $z_{mi},\sigma_i$ and 
the galaxy bias parameter $b_i$ with $i=1,...,10$ and $z_{m1(7)}^c, \sigma_{1(7)}^c, b_{1(7)}^c$ and $f_{1(7)}^c$ 
for the catastrophic sub-samples of bin-$1$($7$), respectively. 
Recall that a fractional $20\%$ prior is applied on $b_i$.  We impose relatively conservative priors on all bins,
\bea
\label{eq:priors}
\nonumber	
\sigma_P(z_{mi})&=&\sigma_i/\sqrt{N_{\rm spec}\times(1-f_i^c)} \\ 
\sigma_P(\sigma_i)&=&\sigma_P(z_{mi})/\sqrt{2} \\ \nonumber
\sigma_P(z_{mi}^c)&=&\sigma_i^c/\sqrt{N_{\rm spec}\times f_i^c} \\ \nonumber
\sigma_P(\sigma_i^c)&=&\sigma_P(z_{mi}^c)/\sqrt{2},
\eea
\noindent by taking $N_{\rm spec}=9$. This translates to 9 sampled spectra in each bin for calibration, 
in the context of Gaussian photo-z errors, i.e., regarding parameters $z_{mi}$ and $\sigma_i$ 
as the photo-z bias (equivalently) and rms errors. Given the Gaussian mean and scatter, 
their errors are subject to a $\sqrt{2}$-times relation \citep[e.g.,][]{ahn03} 
as in the 2nd and 4th lines of equation (\ref{eq:priors}), which we simply take advantage of here.

Provided the survey conditions and fiducial cosmological models, 
the angular power spectra are calculated theoretically using equation (\ref{eq:pl}) and 
are subsequently regarded as the 'observed' data, $\hat{P}_{ij}(\ell)$. 
The covariance matrix, including both shot noise and cosmic variance, is modeled following equation (\ref{eq:cov}).
We next use the CosmoMC package \footnote{http://cosmologist.info/cosmomc/readme.html} \citep{lewis02}, 
with slight modifications to perform the MCMC analysis. We run four chains with about $10^5$ samples 
per chain after burn-in and merge them into one sample. 
The Fisher matrix results are also incorporated as a reference.

\subsection{Results}
\label{sec:mcmc-results}

\begin{figure}
\includegraphics[width=\columnwidth]{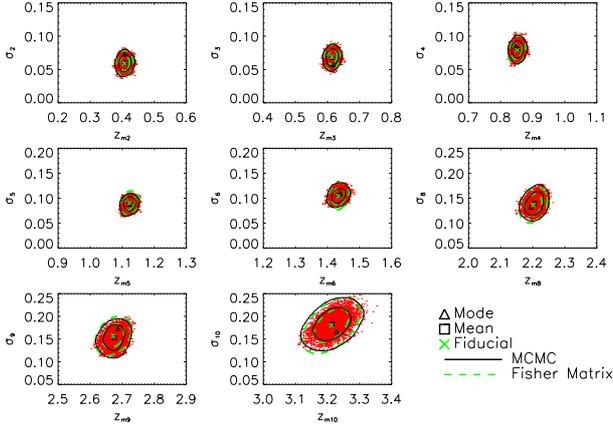}
 \caption{The marginalized constraints on $z_{mi}-\sigma_i$ for bins 
free of catastrophic errors (bin2-6 and 8-10). The black solid contours indicate the
1,2-$\sigma$ credible regions estimated from the MCMC samples
(red dots, thinned to leave $\sim3000$ points for display), with
the mode and mean values marked by triangles and squares,
respectively. The green dashed contours denote predictions by
a Fisher matrix analysis centered at the fiducial values
(green crosses).}
\label{fig:mcmc-otherbins}
\end{figure}

\begin{figure}
\includegraphics[width=\columnwidth]{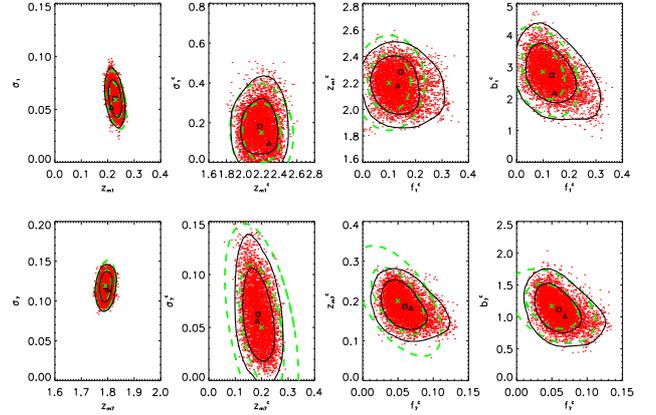}
 \caption{The same as Fig.\ref{fig:mcmc-otherbins} but for parameters of 
bin-1 (upper panels) and bin-7 (lower panels). }
\label{fig:mcmc-2catabins}
\end{figure}

The marginalized constraints on redshift distribution parameters are presented
in Fig.~\ref{fig:mcmc-otherbins} \& \ref{fig:mcmc-2catabins}  for the eight 'no-cata' bins 
and two 'with-cata' bins, respectively. The black solid contours indicate the
1-$\sigma$ and 2-$\sigma$ regions estimated from the MCMC samples
(red dots, thinned to leave $\sim3000$ points for display), with
the mode and mean values marked by triangles and squares,
respectively. The green dashed contours denote predictions by
a Fisher matrix analysis centered at the fiducial values
(green crosses). 

From Fig.~\ref{fig:mcmc-otherbins}, it is evident that for those without-cata bins 
both the best-fit values (mean or mode) of parameters estimated from MCMC and the errors 
are well consistent with the input fiducial values and the error level predicted by 
Fisher matrix analysis, respectively. Roughly speaking, a relative precision 
($\sigma(\theta_i)/\theta_i$) of $\sim1\%-2\%$ on $z_{mi}$ and $\sim 10\%$ on $\sigma_i$ can be achieved. 
The absolute errors in $z_{mi}$ vary from $\sim0.01$ to $\sim0.03$ with increasing $z_{mi}$, 
and $\sim0.007-0.02$ in $\sigma_i$. 

For the 'with-cata' bins shown in Fig.~\ref{fig:mcmc-2catabins}, parameters of 
the main sample distribution of bin-$1$ get less constrained with fractional errors 
$6\%$ on $z_{m1}$ and $17\%$ on $\sigma_{1}$ (absolute errors $\sim0.015$ and $\sim0.011$), 
whereas precision on the corresponding parameters of bin-$7$ are comparable to the other 'no-cata' bins. 
This is mainly due to the high-$\ell$ cut of bin-$1$ discard entirely the BAO signals, 
which can result in a $30-40\%$ loss of precision according to Section \ref{sec:fish-results}. 
Concerning the catastrophic sub-sample in Bin-$1(7)$,  the mean redshift and width, 
$z^c$ and $\sigma^c$, can be reconstructed with a fractional precision $\sim 6(19)\%$ 
and $\sim50(40)\%$ respectively. The catastrophic fraction parameter $f^c$ is constrained 
to a level of $\sim40\%$ for both bins. $f^c$ is considerably degenerated with the galaxy bias parameter $b^c$, 
due to the fact that they both impact the amplitude of the cross power spectrum.  

The MCMC results indicate that it will be feasible in practical analysis to reconstruct the 
true redshift distribution of galaxies with angular power spectra. 
The precision on mean redshift and width of the tomographic bins can generally reach 
the level of $\sigma(z_{mi})/z_{mi}\sim1\%$ and $\sigma(\sigma_i)/\sigma_i\sim10\%$, respectively.
To rewrite the uncertainty in $z_m$ to be $\sigma(z_{mi})/(1+z_{mi})$, which is then roughly equivalent to 
the bias in photo-z errors $e_z=(z_p-z_s)/(1+z_s)$, it gives $\sim0.005-0.008$ for different bins. 
Given the stringent requirement of a bias in $e_z$ below $0.003$ for a LSST-like survey to 
exploit its high statistical precision \citep[e.g.,][]{LSST09}, the spectroscopic sub-sample calibration 
and/or other cross correlation techniques should be combined to bridge the gap to the largest extent.
        
\section{Summary}
\label{sec:sum}

The unprecedented statistical precision of future large weak lensing surveys require 
a rigorous control of systematic effects, such as those of photometric redshifts delivered by multi-band photometry. 
Hence precise calibration of the redshift distribution of the sampled galaxies, $n(z)$, 
is crucial in cosmological parameters estimation.  
The galaxy angular cross power spectra, determined by the true redshift overlap of two tomographic bins, 
are helpful for self-calibrating the photo-z bias and rms errors.

Considering an LSST-like fiducial survey, we investigate the contributions of various components 
of the galaxy angular power spectra in constraining the redshift distribution parameters. 
The principal role of cross spectra is confirmed, whereas the BAO feature is found to be relatively less important in this regard. 

We further investigate, with MCMC analysis, the feasibility of this method in reconstructing 
the underlying redshift distribution when it is confronted with a practical data analysis. 
For the $10$-bin fiducial survey,  we find that the precision on mean redshift and width of 
the tomographic bins can generally reach the level of $1\%$ and $10\%$ ($\sigma(\theta_i)/\theta_i$), respectively. 
For bins with a catastrophic failure fraction, the mean redshift and width of the catastrophic sub-samples, 
$z^c$ and $\sigma^c$, can be constrained to a fractional precision $\sim 5-20\%$ and $\sim40-50\%$, respectively. 
Combined with other calibration methods such as the spectroscopic calibration, 
it is promising to approach the targeted precision on galaxy redshift distribution parameters.

For simplicity, we have ignored the intrinsic and the lensing-induced cross correlations between redshift bins. 
Although the simple double-Gaussian model suffices for this work, 
precision modeling of the underlying redshift distribution is of great importance. 
A principal component analysis directly on the simulated $P(z_s|z_p)$ map may be one of the next directions. 
With a focus on the performance of angular (cross) power spectra in reconstructing the galaxy redshift distributions, 
we fixed the cosmological parameters at their fiducial values. Nevertheless, 
the ultimate goal is to realize the self-calibration, namely, combining the weak lensing and galaxy clustering analysis 
on the very same galaxy sample to constrain both the cosmological and redshift distribution parameters, simultaneously.

\section*{Acknowledgements}

We thank Qiao Wang and Teppei Okumura for useful discussions.
This work was supported by the National Natural Science Foundation of
China grant No. 11033005, the National Key Basic Research Science
Foundation of China grant No. 2010CB833000, the Bairen program from
the Chinese Academy of Sciences, and the Junior Researcher Foundation at
the National Astronomical Observatories of China.




\bibliographystyle{mnras}
\bibliography{bibfile} 

\begin{thebibliography}{}
\makeatletter
\relax
\def\mn@urlcharsother{\let\do\@makeother \do\$\do\&\do\#\do\^\do\_\do\%\do\~}
\def\mn@doi{\begingroup\mn@urlcharsother \@ifnextchar [ {\mn@doi@}
  {\mn@doi@[]}}
\def\mn@doi@[#1]#2{\def\@tempa{#1}\ifx\@tempa\@empty \href
  {http://dx.doi.org/#2} {doi:#2}\else \href {http://dx.doi.org/#2} {#1}\fi
  \endgroup}
\def\mn@eprint#1#2{\mn@eprint@#1:#2::\@nil}
\def\mn@eprint@arXiv#1{\href {http://arxiv.org/abs/#1} {{\tt arXiv:#1}}}
\def\mn@eprint@dblp#1{\href {http://dblp.uni-trier.de/rec/bibtex/#1.xml}
  {dblp:#1}}
\def\mn@eprint@#1:#2:#3:#4\@nil{\def\@tempa {#1}\def\@tempb {#2}\def\@tempc
  {#3}\ifx \@tempc \@empty \let \@tempc \@tempb \let \@tempb \@tempa \fi \ifx
  \@tempb \@empty \def\@tempb {arXiv}\fi \@ifundefined
  {mn@eprint@\@tempb}{\@tempb:\@tempc}{\expandafter \expandafter \csname
  mn@eprint@\@tempb\endcsname \expandafter{\@tempc}}}

\bibitem[\protect\citeauthoryear{{Abdalla}, {Amara}, {Capak}, {Cypriano},
  {Lahav}  \& {Rhodes}}{{Abdalla} et~al.}{2008}]{abdalla08}
{Abdalla} F.~B.,  {Amara} A.,  {Capak} P.,  {Cypriano} E.~S.,  {Lahav} O.,
  {Rhodes} J.,  2008, \mn@doi [\mnras] {10.1111/j.1365-2966.2008.13151.x},
  \href {http://adsabs.harvard.edu/abs/2008MNRAS.387..969A} {387, 969}

\bibitem[\protect\citeauthoryear{Ahn \& Fessler}{Ahn \& Fessler}{2003}]{ahn03}
Ahn S.,  Fessler J.~A.,  2003, EECS Department, The University of Michigan,
  pp~1--2

\bibitem[\protect\citeauthoryear{{Amara} \& {R{\'e}fr{\'e}gier}}{{Amara} \&
  {R{\'e}fr{\'e}gier}}{2007}]{amara07}
{Amara} A.,  {R{\'e}fr{\'e}gier} A.,  2007, \mn@doi [\mnras]
  {10.1111/j.1365-2966.2007.12271.x}, \href
  {http://adsabs.harvard.edu/abs/2007MNRAS.381.1018A} {381, 1018}

\bibitem[\protect\citeauthoryear{{Bacon}, {Refregier}, {Clowe}  \&
  {Ellis}}{{Bacon} et~al.}{2001}]{bacon01}
{Bacon} D.~J.,  {Refregier} A.,  {Clowe} D.,   {Ellis} R.~S.,  2001, \mn@doi
  [\mnras] {10.1046/j.1365-8711.2001.04507.x}, \href
  {http://adsabs.harvard.edu/abs/2001MNRAS.325.1065B} {325, 1065}

\bibitem[\protect\citeauthoryear{{Bartelmann}}{{Bartelmann}}{2010}]{bartelmann%
10}
{Bartelmann} M.,  2010, \mn@doi [Classical and Quantum Gravity]
  {10.1088/0264-9381/27/23/233001}, \href
  {http://adsabs.harvard.edu/abs/2010CQGra..27w3001B} {27, 233001}

\bibitem[\protect\citeauthoryear{{Benjamin}, {Heymans}, {Semboloni}, {van
  Waerbeke}  \& {et al}.}{{Benjamin} et~al.}{2007}]{benjamin07}
{Benjamin} J.,  {Heymans} C.,  {Semboloni} E.,  {van Waerbeke} L.,   {et al}.
  2007, \mn@doi [\mnras] {10.1111/j.1365-2966.2007.12202.x}, \href
  {http://adsabs.harvard.edu/abs/2007MNRAS.381..702B} {381, 702}

\bibitem[\protect\citeauthoryear{{Benjamin}, {van Waerbeke}, {M{\'e}nard}  \&
  {Kilbinger}}{{Benjamin} et~al.}{2010}]{benjamin10}
{Benjamin} J.,  {van Waerbeke} L.,  {M{\'e}nard} B.,   {Kilbinger} M.,  2010,
  \mn@doi [\mnras] {10.1111/j.1365-2966.2010.17191.x}, \href
  {http://adsabs.harvard.edu/abs/2010MNRAS.408.1168B} {408, 1168}

\bibitem[\protect\citeauthoryear{{Bridle} et~al.,}{{Bridle}
  et~al.}{2009}]{bridle09}
{Bridle} S.,  et~al., 2009, \mn@doi [Annals of Applied Statistics]
  {10.1214/08-AOAS222}, \href
  {http://adsabs.harvard.edu/abs/2009AnApS...3....6B} {3, 6}

\bibitem[\protect\citeauthoryear{{Cunha}, {Lima}, {Oyaizu}, {Frieman}  \&
  {Lin}}{{Cunha} et~al.}{2009}]{cunha09}
{Cunha} C.~E.,  {Lima} M.,  {Oyaizu} H.,  {Frieman} J.,   {Lin} H.,  2009,
  \mn@doi [\mnras] {10.1111/j.1365-2966.2009.14908.x}, \href
  {http://adsabs.harvard.edu/abs/2009MNRAS.396.2379C} {396, 2379}

\bibitem[\protect\citeauthoryear{{Eisenstein} \& {Hu}}{{Eisenstein} \&
  {Hu}}{1998}]{eisenstein98}
{Eisenstein} D.~J.,  {Hu} W.,  1998, \mn@doi [\apj] {10.1086/305424}, \href
  {http://adsabs.harvard.edu/abs/1998ApJ...496..605E} {496, 605}

\bibitem[\protect\citeauthoryear{{Eisenstein}, {Seo}  \& {White}}{{Eisenstein}
  et~al.}{2007}]{eisenstein07}
{Eisenstein} D.~J.,  {Seo} H.-J.,   {White} M.,  2007, \mn@doi [\apj]
  {10.1086/518755}, \href {http://adsabs.harvard.edu/abs/2007ApJ...664..660E}
  {664, 660}

\bibitem[\protect\citeauthoryear{{Eriksen} \& {Gazta{\~n}aga}}{{Eriksen} \&
  {Gazta{\~n}aga}}{2015}]{eriksen15}
{Eriksen} M.,  {Gazta{\~n}aga} E.,  2015, \mn@doi [\mnras]
  {10.1093/mnras/stv1288}, \href
  {http://adsabs.harvard.edu/abs/2015MNRAS.452.2149E} {452, 2149}

\bibitem[\protect\citeauthoryear{{Fan}}{{Fan}}{2007}]{fan07}
{Fan} Z.-H.,  2007, \mn@doi [\apj] {10.1086/521182}, \href
  {http://adsabs.harvard.edu/abs/2007ApJ...669...10F} {669, 10}

\bibitem[\protect\citeauthoryear{{Fu} et~al.,}{{Fu} et~al.}{2014}]{fu14}
{Fu} L.,  et~al., 2014, \mn@doi [\mnras] {10.1093/mnras/stu754}, \href
  {http://adsabs.harvard.edu/abs/2014MNRAS.441.2725F} {441, 2725}

\bibitem[\protect\citeauthoryear{Gilks, Richardson  \& Spiegelhalter}{Gilks
  et~al.}{1996}]{gilks96}
Gilks W.~R.,  Richardson S.,   Spiegelhalter D.~J.,  1996, Markov Chain Monte
  Carlo in Practice.
Chapman and Hall, London

\bibitem[\protect\citeauthoryear{{Hearin}, {Zentner}, {Ma}  \&
  {Huterer}}{{Hearin} et~al.}{2010}]{hearin10}
{Hearin} A.~P.,  {Zentner} A.~R.,  {Ma} Z.,   {Huterer} D.,  2010, \mn@doi
  [\apj] {10.1088/0004-637X/720/2/1351}, \href
  {http://adsabs.harvard.edu/abs/2010ApJ...720.1351H} {720, 1351}

\bibitem[\protect\citeauthoryear{{Hearin}, {Zentner}  \& {Ma}}{{Hearin}
  et~al.}{2012}]{hearin12}
{Hearin} A.~P.,  {Zentner} A.~R.,   {Ma} Z.,  2012, \mn@doi [\jcap]
  {10.1088/1475-7516/2012/04/034}, \href
  {http://adsabs.harvard.edu/abs/2012JCAP...04..034H} {4, 34}

\bibitem[\protect\citeauthoryear{{Heymans}, {Brown}, {Heavens}, {Meisenheimer},
  {Taylor}  \& {Wolf}}{{Heymans} et~al.}{2004}]{heymans04}
{Heymans} C.,  {Brown} M.,  {Heavens} A.,  {Meisenheimer} K.,  {Taylor} A.,
  {Wolf} C.,  2004, \mn@doi [\mnras] {10.1111/j.1365-2966.2004.07264.x}, \href
  {http://adsabs.harvard.edu/abs/2004MNRAS.347..895H} {347, 895}

\bibitem[\protect\citeauthoryear{{Heymans} et~al.,}{{Heymans}
  et~al.}{2006}]{heymans06}
{Heymans} C.,  et~al., 2006, \mn@doi [\mnras]
  {10.1111/j.1365-2966.2006.10198.x}, \href
  {http://adsabs.harvard.edu/abs/2006MNRAS.368.1323H} {368, 1323}

\bibitem[\protect\citeauthoryear{{Hoekstra}, {Yee}  \& {Gladders}}{{Hoekstra}
  et~al.}{2002}]{hoekstra02}
{Hoekstra} H.,  {Yee} H.~K.~C.,   {Gladders} M.~D.,  2002, \mn@doi [\apj]
  {10.1086/342120}, \href {http://adsabs.harvard.edu/abs/2002ApJ...577..595H}
  {577, 595}

\bibitem[\protect\citeauthoryear{{Hoekstra}, {Mellier}, {van Waerbeke},
  {Semboloni}  \& {et al}.}{{Hoekstra} et~al.}{2006}]{hoekstra06}
{Hoekstra} H.,  {Mellier} Y.,  {van Waerbeke} L.,  {Semboloni} E.,   {et al}.
  2006, \mn@doi [\apj] {10.1086/503249}, \href
  {http://adsabs.harvard.edu/abs/2006ApJ...647..116H} {647, 116}

\bibitem[\protect\citeauthoryear{{Huterer}, {Takada}, {Bernstein}  \&
  {Jain}}{{Huterer} et~al.}{2006}]{huterer06}
{Huterer} D.,  {Takada} M.,  {Bernstein} G.,   {Jain} B.,  2006, \mn@doi
  [\mnras] {10.1111/j.1365-2966.2005.09782.x}, \href
  {http://adsabs.harvard.edu/abs/2006MNRAS.366..101H} {366, 101}

\bibitem[\protect\citeauthoryear{{Jarvis}, {Bernstein}, {Fischer}, {Smith},
  {Jain}, {Tyson}  \& {Wittman}}{{Jarvis} et~al.}{2003}]{jarvis03}
{Jarvis} M.,  {Bernstein} G.~M.,  {Fischer} P.,  {Smith} D.,  {Jain} B.,
  {Tyson} J.~A.,   {Wittman} D.,  2003, \mn@doi [\aj] {10.1086/367799}, \href
  {http://adsabs.harvard.edu/abs/2003AJ....125.1014J} {125, 1014}

\bibitem[\protect\citeauthoryear{{Jee}, {Tyson}, {Schneider}, {Wittman},
  {Schmidt}  \& {Hilbert}}{{Jee} et~al.}{2013}]{jee13}
{Jee} M.~J.,  {Tyson} J.~A.,  {Schneider} M.~D.,  {Wittman} D.,  {Schmidt} S.,
   {Hilbert} S.,  2013, \mn@doi [\apj] {10.1088/0004-637X/765/1/74}, \href
  {http://adsabs.harvard.edu/abs/2013ApJ...765...74J} {765, 74}

\bibitem[\protect\citeauthoryear{{Joachimi} et~al.,}{{Joachimi}
  et~al.}{2015}]{joachimi15}
{Joachimi} B.,  et~al., 2015, \mn@doi [\ssr] {10.1007/s11214-015-0177-4}, \href
  {http://adsabs.harvard.edu/abs/2015SSRv..tmp...65J} {}

\bibitem[\protect\citeauthoryear{{Kilbinger}}{{Kilbinger}}{2015}]{kilbinger15}
{Kilbinger} M.,  2015, \mn@doi [Reports on Progress in Physics]
  {10.1088/0034-4885/78/8/086901}, \href
  {http://adsabs.harvard.edu/abs/2015RPPh...78h6901K} {78, 086901}

\bibitem[\protect\citeauthoryear{{Kilbinger}, {Fu}, {Heymans}, {Simpson}  \&
  {et al}.}{{Kilbinger} et~al.}{2013}]{kilbinger13}
{Kilbinger} M.,  {Fu} L.,  {Heymans} C.,  {Simpson} F.,   {et al}. 2013,
  \mn@doi [\mnras] {10.1093/mnras/stt041}, \href
  {http://adsabs.harvard.edu/abs/2013MNRAS.430.2200K} {430, 2200}

\bibitem[\protect\citeauthoryear{{K{\"o}hlinger}, {Viola}, {Valkenburg},
  {Joachimi}, {Hoekstra}  \& {Kuijken}}{{K{\"o}hlinger}
  et~al.}{2015}]{Kohlinger15}
{K{\"o}hlinger} F.,  {Viola} M.,  {Valkenburg} W.,  {Joachimi} B.,  {Hoekstra}
  H.,   {Kuijken} K.,  2015, preprint, \href
  {http://adsabs.harvard.edu/abs/2015arXiv150904071K} {} (\mn@eprint {arXiv}
  {1509.04071})

\bibitem[\protect\citeauthoryear{{LSST Science Collaboration} et~al.,}{{LSST
  Science Collaboration} et~al.}{2009}]{LSST09}
{LSST Science Collaboration} et~al., 2009, preprint, \href
  {http://adsabs.harvard.edu/abs/2009arXiv0912.0201L} {} (\mn@eprint {arXiv}
  {0912.0201})

\bibitem[\protect\citeauthoryear{{Lewis} \& {Bridle}}{{Lewis} \&
  {Bridle}}{2002}]{lewis02}
{Lewis} A.,  {Bridle} S.,  2002, \mn@doi [\prd] {10.1103/PhysRevD.66.103511},
  \href {http://adsabs.harvard.edu/abs/2002PhRvD..66j3511L} {66, 103511}

\bibitem[\protect\citeauthoryear{{Lima}, {Cunha}, {Oyaizu}, {Frieman}, {Lin}
  \& {Sheldon}}{{Lima} et~al.}{2008}]{lima08}
{Lima} M.,  {Cunha} C.~E.,  {Oyaizu} H.,  {Frieman} J.,  {Lin} H.,   {Sheldon}
  E.~S.,  2008, \mn@doi [\mnras] {10.1111/j.1365-2966.2008.13510.x}, \href
  {http://adsabs.harvard.edu/abs/2008MNRAS.390..118L} {390, 118}

\bibitem[\protect\citeauthoryear{{Liu} et~al.,}{{Liu} et~al.}{2015}]{liu14}
{Liu} X.,  et~al., 2015, \mn@doi [\mnras] {10.1093/mnras/stv784}, \href
  {http://adsabs.harvard.edu/abs/2015MNRAS.450.2888L} {450, 2888}

\bibitem[\protect\citeauthoryear{{Ma}, {Hu}  \& {Huterer}}{{Ma}
  et~al.}{2006}]{ma06}
{Ma} Z.,  {Hu} W.,   {Huterer} D.,  2006, \mn@doi [\apj] {10.1086/497068},
  \href {http://adsabs.harvard.edu/abs/2006ApJ...636...21M} {636, 21}

\bibitem[\protect\citeauthoryear{{Massey}, {Rhodes}, {Leauthaud}, {Capak}  \&
  {et al}.}{{Massey} et~al.}{2007}]{massey07}
{Massey} R.,  {Rhodes} J.,  {Leauthaud} A.,  {Capak} P.,   {et al}. 2007,
  \mn@doi [\apjs] {10.1086/516599}, \href
  {http://adsabs.harvard.edu/abs/2007ApJS..172..239M} {172, 239}

\bibitem[\protect\citeauthoryear{{Matthews} \& {Newman}}{{Matthews} \&
  {Newman}}{2010}]{matthews10}
{Matthews} D.~J.,  {Newman} J.~A.,  2010, \mn@doi [\apj]
  {10.1088/0004-637X/721/1/456}, \href
  {http://adsabs.harvard.edu/abs/2010ApJ...721..456M} {721, 456}

\bibitem[\protect\citeauthoryear{{McQuinn} \& {White}}{{McQuinn} \&
  {White}}{2013}]{mcquinn13}
{McQuinn} M.,  {White} M.,  2013, \mn@doi [\mnras] {10.1093/mnras/stt914},
  \href {http://adsabs.harvard.edu/abs/2013MNRAS.433.2857M} {433, 2857}

\bibitem[\protect\citeauthoryear{{M{\'e}nard}, {Scranton}, {Schmidt},
  {Morrison}, {Jeong}, {Budavari}  \& {Rahman}}{{M{\'e}nard}
  et~al.}{2013}]{menard13}
{M{\'e}nard} B.,  {Scranton} R.,  {Schmidt} S.,  {Morrison} C.,  {Jeong} D.,
  {Budavari} T.,   {Rahman} M.,  2013, preprint, \href
  {http://adsabs.harvard.edu/abs/2013arXiv1303.4722M} {} (\mn@eprint {arXiv}
  {1303.4722})

\bibitem[\protect\citeauthoryear{{Miller}, {Kitching}, {Heymans}, {Heavens}  \&
  {van Waerbeke}}{{Miller} et~al.}{2007}]{miller07}
{Miller} L.,  {Kitching} T.~D.,  {Heymans} C.,  {Heavens} A.~F.,   {van
  Waerbeke} L.,  2007, \mn@doi [\mnras] {10.1111/j.1365-2966.2007.12363.x},
  \href {http://adsabs.harvard.edu/abs/2007MNRAS.382..315M} {382, 315}

\bibitem[\protect\citeauthoryear{{Rahman}, {M{\'e}nard}, {Scranton}, {Schmidt}
  \& {Morrison}}{{Rahman} et~al.}{2015}]{rahman15}
{Rahman} M.,  {M{\'e}nard} B.,  {Scranton} R.,  {Schmidt} S.~J.,   {Morrison}
  C.~B.,  2015, \mn@doi [\mnras] {10.1093/mnras/stu2636}, \href
  {http://adsabs.harvard.edu/abs/2015MNRAS.447.3500R} {447, 3500}

\bibitem[\protect\citeauthoryear{{Refregier}}{{Refregier}}{2003}]{refregier03}
{Refregier} A.,  2003, \mn@doi [\araa]
  {10.1146/annurev.astro.41.111302.102207}, \href
  {http://adsabs.harvard.edu/abs/2003ARA%26A..41..645R} {41, 645}

\bibitem[\protect\citeauthoryear{{Rhodes}, {Refregier}, {Collins}, {Gardner},
  {Groth}  \& {Hill}}{{Rhodes} et~al.}{2004}]{rhodes04}
{Rhodes} J.,  {Refregier} A.,  {Collins} N.~R.,  {Gardner} J.~P.,  {Groth}
  E.~J.,   {Hill} R.~S.,  2004, \mn@doi [\apj] {10.1086/382181}, \href
  {http://adsabs.harvard.edu/abs/2004ApJ...605...29R} {605, 29}

\bibitem[\protect\citeauthoryear{{Schmidt}, {M{\'e}nard}, {Scranton},
  {Morrison}  \& {McBride}}{{Schmidt} et~al.}{2013}]{schmidt13}
{Schmidt} S.~J.,  {M{\'e}nard} B.,  {Scranton} R.,  {Morrison} C.,   {McBride}
  C.~K.,  2013, \mn@doi [\mnras] {10.1093/mnras/stt410}, \href
  {http://adsabs.harvard.edu/abs/2013MNRAS.431.3307S} {431, 3307}

\bibitem[\protect\citeauthoryear{{Schneider}, {Knox}, {Zhan}  \&
  {Connolly}}{{Schneider} et~al.}{2006}]{schneider06}
{Schneider} M.,  {Knox} L.,  {Zhan} H.,   {Connolly} A.,  2006, \mn@doi [\apj]
  {10.1086/507675}, \href {http://adsabs.harvard.edu/abs/2006ApJ...651...14S}
  {651, 14}

\bibitem[\protect\citeauthoryear{{Semboloni}, {Mellier}, {van Waerbeke},
  {Hoekstra}  \& {et al}.}{{Semboloni} et~al.}{2006}]{semboloni06}
{Semboloni} E.,  {Mellier} Y.,  {van Waerbeke} L.,  {Hoekstra} H.,   {et al}.
  2006, \mn@doi [\aap] {10.1051/0004-6361:20054479}, \href
  {http://adsabs.harvard.edu/abs/2006A%26A...452...51S} {452, 51}

\bibitem[\protect\citeauthoryear{{Sun}, {Fan}, {Tao}, {Kneib}, {Jouvel}  \&
  {Tilquin}}{{Sun} et~al.}{2009}]{sun09}
{Sun} L.,  {Fan} Z.-H.,  {Tao} C.,  {Kneib} J.-P.,  {Jouvel} S.,   {Tilquin}
  A.,  2009, \mn@doi [\apj] {10.1088/0004-637X/699/2/958}, \href
  {http://adsabs.harvard.edu/abs/2009ApJ...699..958S} {699, 958}

\bibitem[\protect\citeauthoryear{{Sun}, {Wang}  \& {Zhan}}{{Sun}
  et~al.}{2013}]{sun13}
{Sun} L.,  {Wang} Q.,   {Zhan} H.,  2013, \mn@doi [\apj]
  {10.1088/0004-637X/777/1/75}, \href
  {http://adsabs.harvard.edu/abs/2013ApJ...777...75S} {777, 75}

\bibitem[\protect\citeauthoryear{{Tegmark}}{{Tegmark}}{1997}]{tegmark97}
{Tegmark} M.,  1997, \mn@doi [Physical Review Letters]
  {10.1103/PhysRevLett.79.3806}, \href
  {http://adsabs.harvard.edu/abs/1997PhRvL..79.3806T} {79, 3806}

\bibitem[\protect\citeauthoryear{{Yu}, {Zhang}, {Lin}  \& {Cui}}{{Yu}
  et~al.}{2015}]{yu15}
{Yu} Y.,  {Zhang} P.,  {Lin} W.,   {Cui} W.,  2015, \mn@doi [\apj]
  {10.1088/0004-637X/803/1/46}, \href
  {http://adsabs.harvard.edu/abs/2015ApJ...803...46Y} {803, 46}

\bibitem[\protect\citeauthoryear{{Zhan}}{{Zhan}}{2006}]{zhan06}
{Zhan} H.,  2006, \mn@doi [\jcap] {10.1088/1475-7516/2006/08/008}, \href
  {http://adsabs.harvard.edu/abs/2006JCAP...08..008Z} {8, 8}

\bibitem[\protect\citeauthoryear{{Zhan} \& {Knox}}{{Zhan} \&
  {Knox}}{2006}]{zhan06b}
{Zhan} H.,  {Knox} L.,  2006, \mn@doi [\apj] {10.1086/503622}, \href
  {http://adsabs.harvard.edu/abs/2006ApJ...644..663Z} {644, 663}

\bibitem[\protect\citeauthoryear{{Zhan}, {Knox}, {Tyson}  \&
  {Margoniner}}{{Zhan} et~al.}{2006}]{Zhan06a}
{Zhan} H.,  {Knox} L.,  {Tyson} J.~A.,   {Margoniner} V.,  2006, \mn@doi [\apj]
  {10.1086/500077}, \href {http://adsabs.harvard.edu/abs/2006ApJ...640....8Z}
  {640, 8}

\bibitem[\protect\citeauthoryear{{Zhang}, {Pen}  \& {Bernstein}}{{Zhang}
  et~al.}{2010}]{zhang10}
{Zhang} P.,  {Pen} U.-L.,   {Bernstein} G.,  2010, \mn@doi [\mnras]
  {10.1111/j.1365-2966.2010.16445.x}, \href
  {http://adsabs.harvard.edu/abs/2010MNRAS.405..359Z} {405, 359}

\makeatother
\end{thebibliography}


\bsp	
\label{lastpage}
\end{document}